# Accepted Manuscript

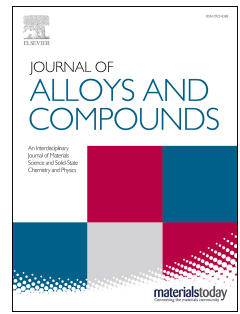

Stable, carbon-free inks of Cu$_2$ZnSnS$_4$ nanoparticles synthesized at room temperature designed for roll-to-roll fabrication of solar cell absorber layers

C. Rein, S. Engberg, J.W. Andreasen

Please cite this article as: C. Rein, S. Engberg, J.W. Andreasen, Stable, carbon-free inks of Cu$_2$ZnSnS$_4$ nanoparticles synthesized at room temperature designed for roll-to-roll fabrication of solar cell absorber layers, *Journal of Alloys and Compounds* (2019), doi: https://doi.org/10.1016/j.jallcom.2019.02.014.

Stable, carbon-free inks of $Cu_2ZnSnS_4$ nanoparticles synthesized at room temperature designed for roll-to-roll fabrication of solar cell absorber layers

C. Rein,[a] S. Engberg[b] and J. W. Andreasen[a]

a. Technical University of Denmark, Department of Energy Conversion and Storage, Frederiksborgvej 399, 4000 Roskilde, Denmark.

b. Technical University of Denmark, Department of Photonics Engineering, Frederiksborgvej 399, 4000 Roskilde, Denmark.

Abstract:

We report on a novel room temperature approach for the synthesis of environmentally-friendly copper zinc tin sulfide ($Cu_2ZnSnS_4$) nanoparticles. The method is shown to be compositionally robust and able to produce $S^{2-}$-stabilized carbon-free nanoparticle inks that are suitable for an absorber layer in solar cells. No organic residues from the process were detected. The metal-composition and the occurrence of secondary phases is here correlated with synthesis conditions: By utilizing a reactant concentration of Cu/Sn < 1.8 and Sn(II) as tin-source it is possible to avoid the formation of $Cu_xS$-phases, which are detrimental for the solar cell performance when present in the final absorber layer. With nanoparticle sizes approaching the Bohr radius for $Cu_2ZnSnS_4$, the band gap can be broadened up to 1.7 eV. In addition, the conditions for forming stable, carbon-free aqueous inks of such $Cu_2ZnSnS_4$ nanoparticles are investigated and the stabilizing $NH_4^+/S^{2-}$-ion concentration affects the quality of the deposited absorber layer. The use of room temperature synthesis and stable aqueous ink formulations make the method suitable for roll-to-roll fabrication and upscaling.

### 1. Introduction

It is becoming increasingly clear that it will be necessary to realize the transition to sustainable, climate friendly energy sources at a very high rate. Amongst other energy technologies, this calls for solar energy with very low Energy Pay-Back Time (EPBT) and with minimal impact on the environment and resources, thus relying on non-toxic and abundant materials[1].

Earth-abundant and inexpensive copper-zinc-tin-sulfide materials (CZTS with standard stoichiometry of $Cu_2ZnSnS_4$) can be used as efficient photo-absorbers for 3rd generation solar cells[2,3]. The favorable direct band gap photovoltaic properties and cheap synthesis of thermodynamically stable CZTS makes it a strong candidate for supporting humanity's growing energy demands in a cost effective and environmentally benign way[4]. Synthesis of CZTS nanoparticles (NP) is often performed at minimum 200 °C in order to activate the monomer precursors and to produce the material without detrimental secondary phases, such as $Cu_xS$ or $SnS$[5,6]. Even at these elevated temperatures, the stoichiometric ratio between the three metals in the reaction medium still needs to be carefully controlled because of the narrow region in the phase space where CZTS can form[7]. The quality of the absorber material, based on synthesized CZTS NPs, can further be improved by annealing at temperatures around 550 °C in sulfur atmosphere. The need for high temperature processing has a significant negative effect on the EPBT.

Upscaling CZTS production to meet the global terawatt consumption requires robust fabrication processes that are compatible with existing roll-to-roll (R2R) manufacturing, without introducing energy-costly steps[8]. As inexpensive, R2R-compatible substrates melt at temperatures above 120 °C this excludes most methods for direct R2R-synthesis of CZTS. Low temperature synthesis of colloidal CZTS has previously been demonstrated[9], but ligands are known to be detrimental for the final absorber layer in the solar cell[8]. Room temperature synthesis of CZTS suitable for R2R absorber fabrication has been demonstrated with both successive ionic layer adsorption and reaction (SILAR)[10] and instantaneous reactions[11]. These methods, however, produce either amorphous material or binary sulfide by-products that are detrimental for solar cell efficiency and require removal by post-synthesis annealing. It is therefore important to improve these methods to make CZTS suitable as absorber material in solar cells, while avoiding the use of high temperature post-processing. Recently, ligand-exchange has been utilized to form absorber layers of electronically connected and completely passivated CZTS, which could potentially be used on an R2R-platform[12], but the need for high temperatures (280 °C) to produce the material still affects the final EPBT negatively.

The preferred delivery method of CZTS NPs on R2R compatible substrates is via concentrated inks usable for printing or spin-coating. Obtaining even a semi-stable concentrated ink of NPs requires an effective passivation of the NP surfaces and a strong NP-to-solvent interaction. This has previously been obtained by using functionalized alkane ligands to coat the NP surface, which would provide a favorable interaction with most organic solvents. Unfortunately, the addition of long hydrocarbons results in the detrimental carbon contaminations in the CZTS absorber layer after deposition, and it is thus interesting to find carbon-free ink formulations. Several groups have already identified various ions[12-14] and complexes[15] that can be used to stabilize CZTS NP inks in hydrophilic solvents. The effectiveness of this stabilization depends on the charge of these ions and complexes, which is determined by the pH of the solutions[16].

Here, we present a room temperature synthesis that produces carbon-free CZTS NPs while avoiding the detrimental binary phases, SnS and $Cu_xS$, and the organic residues from the synthesis. The 6 nm NPs have a copper-poor and zinc-rich composition, optimal for CZTS solar cells[17], with Cu/(Zn + Sn) = 0.8 and Zn/Sn = 1.2-1.3. In addition, we present carbon-free aqueous inks of CZTS NP suitable for the R2R-platform.

The quality of the synthesized CZTS NPs was found to vary depending on the metal-ratios and reaction times. An upper threshold at a reactant Cu/Sn-ratio of 1.7 - 1.8 was identified, above which the detrimental $Cu_xS$ secondary phases form. Additionally, ink stabilization by $(Sn_2S_6)^{4-}$, $S^{2-}$ and $NH_4^+$ was utilized to prepare stable aqueous inks, with $(Sn_2S_6)^{4-}$ later being exchanged with ethylendiamine (EDA) as a "linker" to agglomerate CZTS NPs and destabilize the inks. $S^{2-}$ and $NH_4^+$ concentrations were furthermore found to affect the prevalence of $Cu_xS$-phases, with high $S^{2-}$ concentration completely eliminating this phase.

## 2. Experimental section
### 2.1. Cu₂ZnSnS₄ synthesis

As in a previously reported protocol for synthesizing CZTS NPs[11], two deoxygenated precursor solutions were prepared in advance, and the reaction took place in a nitrogen glove box (Cleaver Scientific, UK). In contrast to the previously reported procedure the Sn-source was changed from $SnCl_4$ to $SnCl_2$ (less toxic) and the reactant Cu/Sn-ratio was reduced from 1.9 to 1.7 - 1.8 (observed threshold for $Cu_xS$ formation). The reactant Zn/Sn- and Cu/Sn-ratios were varied between 0.63 - 1.20 and 1.0 - 1.8, respectively. The samples presented in this paper were obtained from a Cu/Sn-ratio of 1.7 (Zn/Sn=1.09) and 1.8 (Zn/Sn=1.13) and labeled "Low-Cu" and "High-Cu", respectively. Extractions were taken after 5, 15 and 25 min. reaction time.

Precursor A: In a 250 ml round bottom flask 3.61-3.91 mmol of CuCl (99.995%, Sigma-Aldrich), 2.30-3.03 mmol of $ZnCl_2$ (99.999%, Sigma-Aldrich) and 2.11-3.76 mmol $SnCl_2$ (99.99%, Sigma-Aldrich) were first added to 10 ml degassed (multiple vacuum/$N_2$-cycles for 30 min.) acetonitrile and mixed under magnetic stirring at room temperature for minimum 10 min. to form a clear solution (~1 M concentration in total metal cations).

Precursor B: Next, 11.1 mmol of NaSH · 2H₂O (Sigma-Aldrich) was added to 100 ml degassed (multiple vacuum/$N_2$-cycles for 30 min.) water (milli-q grade) and mixed under magnetic stirring at room temperature for minimum 10 min.

Reaction and washing: 10 mL of precursor A was injected into precursor B, causing the solution in the flask to turn black immediately. The NaSH solution was in 30 to 52% molar excess with respect to metal stoichiometry. The solution was allowed to react for 25 min. with small samples extracted at 5 and 15 min. after the injection of precursor A. After reaction, both samples and the final solution were quenched in water and centrifuged at 6,000 rpm for 10 min. to obtain a clear reddish-brown supernatant and a black solid. The supernatant was discarded and the black solid was redispersed in water (approx. 10 min. sonication) before centrifugation. The washing steps were repeated using ethanol to remove unreacted precursors and impurities from the product. Finally, the product was dried in air at 80 °C for 20 hours to ensure the removal of trace acetonitrile ($T_b$ = 82 °C).

### 2.2. NP characterization

The synthesized particles were suspended in ethanol and drop-casted onto a silicon or glass substrate. Energy dispersive X-ray spectroscopy (EDX) was performed using a tabletop scanning electron microscope (SEM) TM3000 (Hitachi) with an accelerating voltage of 15 kV with a Bruker Quantax 70 system. Copper and zinc concentrations were derived from K-transitions for the metals, whereas tin concentration was derived from the fluorescence peaks belonging to the L-transition. Transmission electron microscopy (TEM) and high-resolution TEM (HRTEM) were performed using a JEOL 2100 TEM with an accelerating voltage of 200 kV.

X-ray diffraction (XRD) was carried out with a Bruker D8 Advance X-ray diffractometer with Cu Kα radiation (λ = 0.15418 nm).

Visible and near-UV Raman spectra were acquired using a Renishaw InVia spectrometer with a 532 nm and 325 nm laser, respectively. Each sample spectrum used for analysis was obtained by averaging multiple spectra taken from different areas on the same sample. Analysis of visible light Raman spectra used Gaussian fits for the CZTS-peak range (300-370 $cm^{-1}$) and the $Cu_xS$-peak range (450-500 $cm^{-1}$) with a linear background level. The reference signal from the substrate ($SiO_2$) was subtracted from the near-UV Raman spectra.

Optical diffuse reflectance measurements were carried out with an integrating sphere setup (reflectance geometry 8/d), using a laser-driven white light source (LDLS ENERGETIQ, 170-2000 nm) which was coupled to the sphere using a 600 μm optical fiber. The reflected light was collected with a fiber optics spectrophotometer (QE65000 from Ocean Optics). All reflectance measurements were calibrated with a diffuse reflectance standard and an optical trap. The diffuse reflectance data were converted to the equivalent absorption using the Kulbeka-Munk model, from which the optical band gap was extracted.

Fourier Transform Infra Red spectroscopy (FTIR) was performed using a Tensor 27 from Bruker together with the OPUS software package. The sample stage was cleaned with ethanol between each measurement. Each measurement consisted of 32 scans with 2 $cm^{-1}$ resolution. Data was baseline corrected using a 1$^{st}$ order polynomial, but not rescaled.

### 2.3. Ink formulations and spin-coating

Inks of 10 mg/ml were formed by dispersing the Low-Cu CZTS NPs in milli-q quality water containing 1-100 mM NaSH and 0.01-10 M ($NH_4$)OH. The inks were sonicated for two hours and a few drops were spin-coated on clean glass substrates for Raman spectroscopy (section 2.3), while the remaining inks were left standing for 20 hours for stability investigation.

### 3. Results and discussion

Replicating the method reported by Larramona *et al.*[11] using $SnCl_4$ as Sn-source with a reactant Cu:Zn:Sn composition of 1.9:1.2:1.0 in a 5 min. reaction, we observe the formation of significant amounts of $Cu_xS$-phases, that may be detrimental depending on application, as described in the paper[11]. To redissolve such phases, more than a single high temperature-annealing step or treatment with toxic KCN (Fig. S1, ESI) are needed, but for up-scaled CZTS production, it is desirable to avoid such treatments in the first place. Thus the room temperature synthesis of CZTS NPs for low-cost ink production was challenged by cost (double annealing), quality ($Cu_xS$-phases) and toxicity (KCN and $SnCl_4$). An interesting reaction kinetics suggested by Larramona *et al.*[11] is the possible oxidation of $Sn^{2+}$ by $H_2S$ during the reaction to form CZTS, while consuming the binary copper phase. For our investigations, we therefore replaced $SnCl_4$ with the less toxic $SnCl_2$, but using a reactant Cu:Zn:Sn composition of 1.8:1.1:1.0, some $Cu_xS$-phases were still observed in the final product. Thus, reactant composition had to be adjusted until Cu by-products were no longer detected with Raman spectroscopy.

Five different reactant compositions were tested in total (Fig. S2, ESI), and High-Cu and Low-Cu represent two sides of a threshold above which Cu by-products form. Average Cu content relative to Zn and Sn in the products was 5% lower compared to the reactant concentration. It is difficult to ascertain any trends in compositional dynamics due to uncertainty, but average Cu/(Zn+Sn) appear decreasing for High-Cu and increasing with half the rate for Low-Cu.

The Cu/(Zn+Sn) and Zn/Sn-ratios in the synthesized products are compared to the reactant ratios for each reaction (Fig. 1). Metal ratios were found to change only slightly over time with a constant excess of tin, indicating either the presence of $SnS_2$ or Sn-rich CZTS. When treated with $(NH_4)_2S$, $SnS_2$ can be dissolved in water and form $(NH_3)_2Sn_2S_6$ (aq) that helps stabilize CZTS NPs.[15] Thiostannate complexes can form in a broad range of concentrations and pH levels, with $(Sn_2S_6)^{4-}$ mainly forming in alkaline solutions.[16]

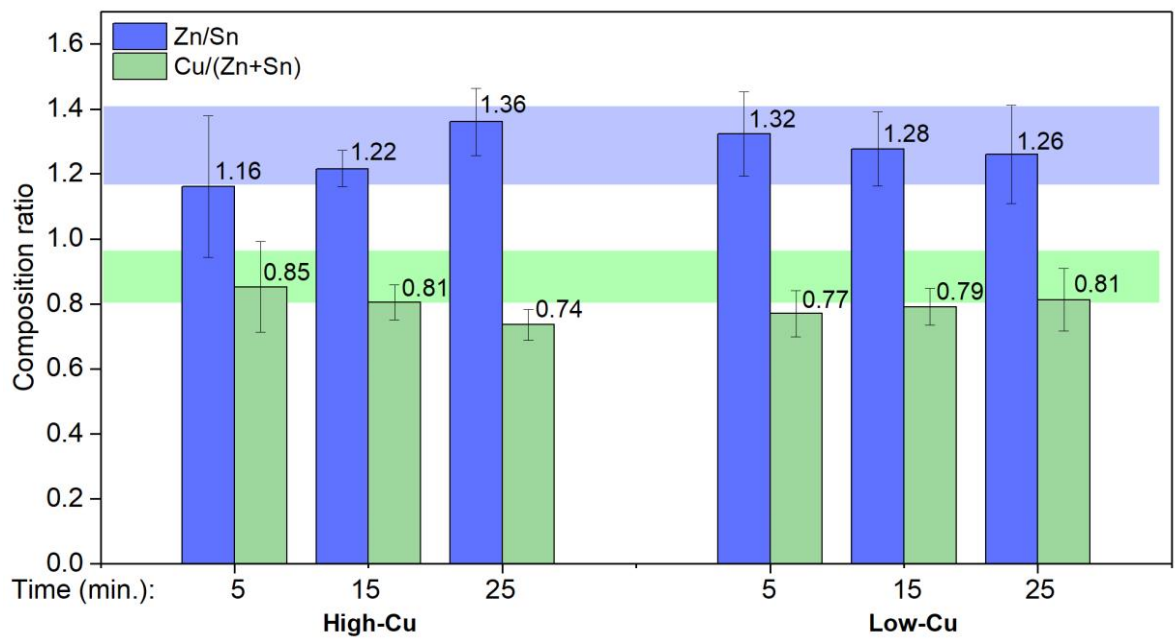

**Fig. 1** Average composition of High-Cu and Low-Cu NPs after 15 min. reaction measured by EDX. Blue and green regions mark the desired Zn/Sn- and Cu/(Zn+Sn)-composition for photovoltaic applications, which are 1.1 - 1.3 and 0.76 - 0.9, respectively.[18]

Both Zn/Sn- and Cu/(Zn+Sn)-ratios in the products correlate well with the reactant ratios, but with some dynamical trends. The largest variation in product composition is observed within the first 5 min. of preparing High-Cu, indicating an initial uneven distribution of elements in the reaction medium. Minimal variation in product composition for all samples (Fig. S2, ESI) was found after 15 min. reaction time suggesting it to be the minimum time needed for a homogenized synthesis.

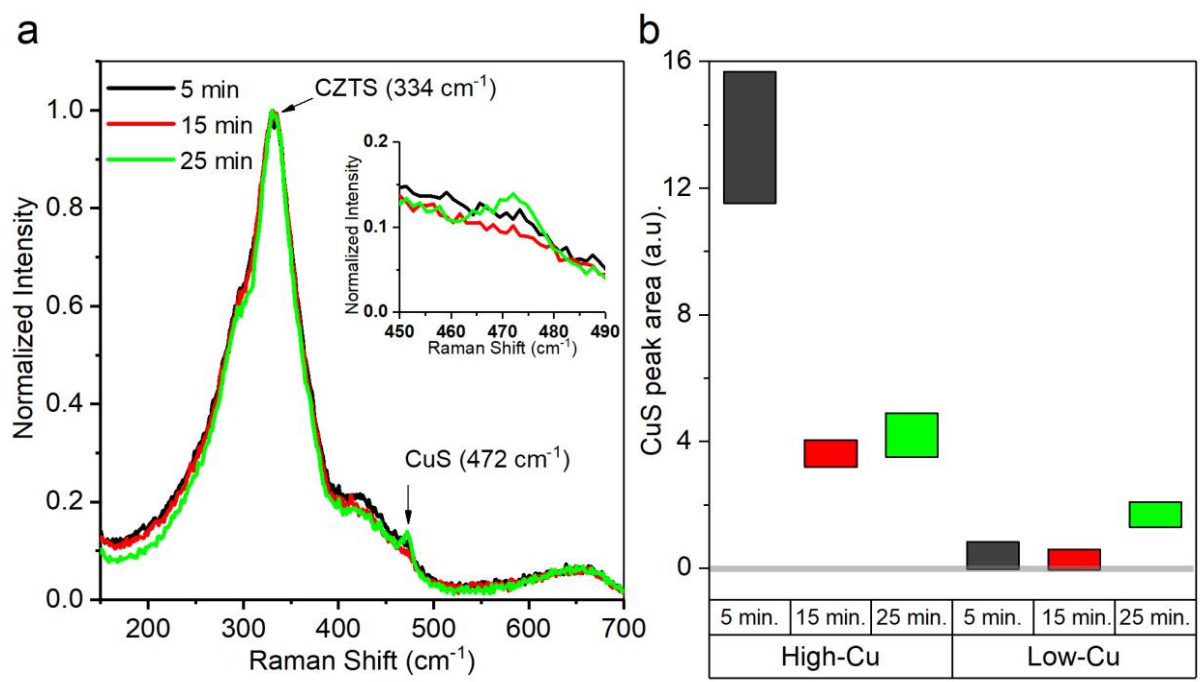

**Fig. 2** (a) 532 nm Raman scattering data for Low-Cu after 5, 15 and 25 min. reaction time. Approximate positions of peaks for CZTS and $Cu_xS$ are shown (CuS as reference). (b) Area and 95% confidence bands of Gaussian fitted peaks at 472 cm$^{-1}$ (Fig. S3, ESI). The confidence band is required to have an entirely positive peak area in order to prove the presence of $Cu_xS$. We see no $Cu_xS$ in the 5 and 15 min. reaction times of Low-Cu.

Raman spectroscopy reveals a clear and persistent presence of kesterite CZTS based on the primary A-mode peak in all samples (334 cm$^{-1}$, Fig. 2a). The peak is slightly shifted and broadened compared to previous observations for kesterite CZTS NP (336 cm$^{-1}$)[6], which is due to the decreased correlation length (phonon confinement) in small NPs[19]. The second A-mode related to kesterite-structured CZTS at approximately 290 cm$^{-1}$ is also visible as a shoulder due to peak-broadening. We notice a small peak at approx. 472 cm$^{-1}$ for 25 min. reaction time (Fig. 2a, insert), which suggests the presence of $Cu_xS$ in the sample and going from Low-Cu to High-Cu leads to a significant increase in $Cu_xS$-related peak intensity (Fig. 2b). Copper sulfide binary phases have characteristic Raman shifts in this region and, when present, the low concentration of these phases makes the precise identification of phases difficult. We therefore use $Cu_xS$ as a collective term for these detrimental phases. The 15 min. reaction appears to produce the lowest $Cu_xS$ content in all samples indicating this reaction time to be optimal for the synthesis conditions. No clear correlation between EDX-derived Cu content and the $Cu_xS$/CZTS-values obtained by Raman spectroscopy is observed, which suggests that using Cu/(Zn+Sn) = 0.82 in the reaction facilitates the highest Cu content in the product without forming any $Cu_xS$-phases.

No triple peaks at 160, 190 and 220 cm$^{-1}$ were observed for any sample, indicating the absence of SnS, whereas the broad CZTS peak might hide the signal from any possible $SnS_2$-phase (314 cm$^{-1}$)[20]. The lack of SnS could indicate oxidation of most $Sn^{2+}$ ions to $Sn^{4+}$ by $H_2S$ as described previously.

The high Zn/Sn ratio was expected to result in the formation of ZnS, where the high band gap (3.5 eV) material will act as non-functional area in the absorber layer without decreasing the efficiency of the remaining CZTS in the layer. For further identification of binary sulfides such as $Cu_xS$ and especially ZnS, near-UV Raman spectroscopy was used because it resonates with the band gap of ZnS and suppresses the broad CZTS-related signals. No $Cu_xS$-phases and only trace amounts of ZnS could be observed (Fig. 3). The little/no trace of ZnS by-product correlates well with previous findings, where low temperature synthesis was found to promote the incorporation of Zn into CZTS and inhibit ZnS-formation[21]. Such incorporation would promote the formation of charge neutral [$V_{Cu}$ + $Zn_{Cu}$] defect pairs that easily forms in CZTS, and which can be beneficial for the photovoltaic efficiency of CZTS[22]. In addition, the lack of ZnS signal could also be due to ZnS-clusters being smaller than the detectable limit.

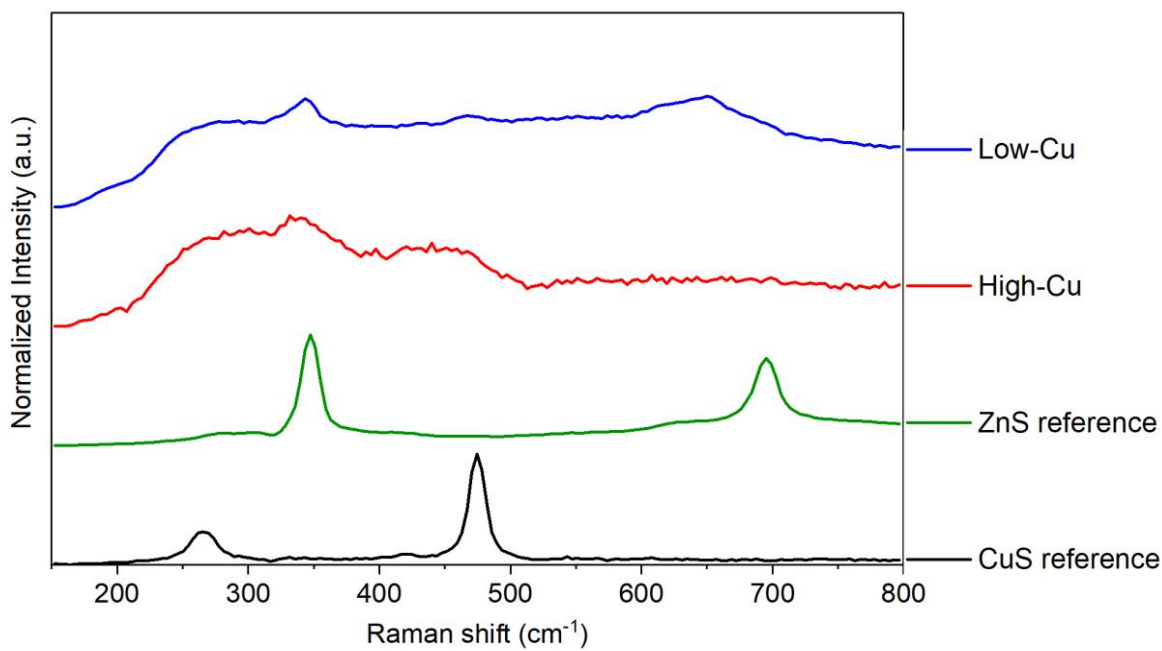

**Fig. 3** Background-subtracted near-UV Raman data of 15 min. samples and reference binary sulfide compounds. Only one sample (Cu/Sn = 1.0) has a peak near 475 cm$^{-1}$ which indicates a presence of Cu$_x$S.

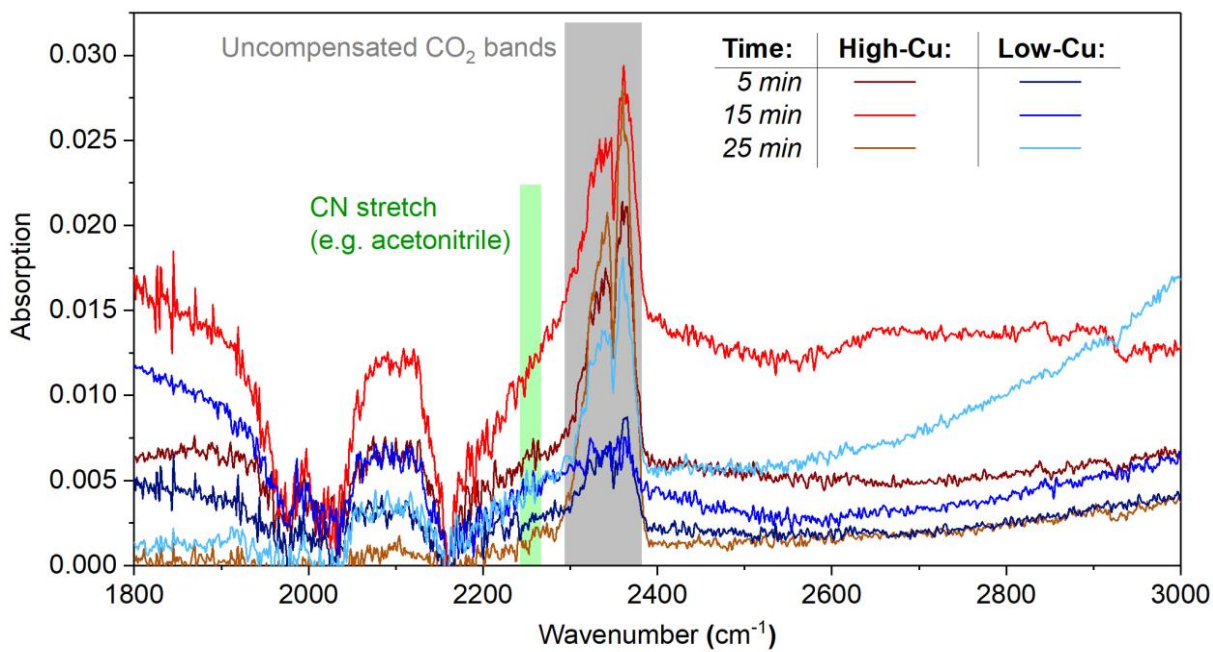

**Fig. 4:** FTIR spectra of High-Cu (red) and Low-Cu (blue) CZTS NPs. Acetonitrile has a strong CN stretch peak around 2267 cm$^{-1}$ (green region), and the lack of this peak indicates that there is no acetonitrile residues remaining from the synthesis. Uncompensated CO$_2$ bands are found in the 2290-2390 cm$^{-1}$ region (grey region).

FTIR of the products was used to verify that no acetonitrile residues remained on the surface of the CZTS NPs after several washing steps and overnight drying (Fig. 4). FTIR allows us to easily detect the functional groups of organic molecules, and acetonitrile contain a CN-group that gives rise to a strong absorption peak at 2267 cm$^{-1}$. This value can shift slightly depending on the environment around the molecule and a peak in the 2250-2280 cm$^{-1}$ region would indicate the presence of acetonitrile: This region is highlighted as a green region in Fig. 4. No peaks are seen and thus no organic ligands coat the surface of the CZTS NPs. The full range FTIR spectra without baseline corrections can be found in the ESI (Fig. S4, ESI).

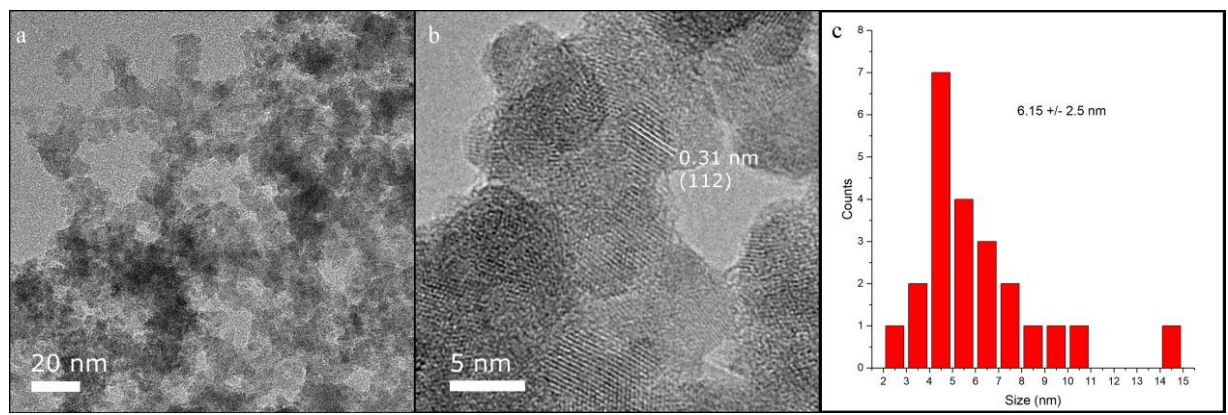

**Fig. 5** TEM and HRTEM analysis of Low-Cu NPs after 15 min. reaction time showing (a) low resolution TEM image, (b) inter-planar spacing of lattice fringes from different planes observed in distinct CZTS NPs, and (c) a histogram size analysis of the CZTS NPs shown in (a).

The morphology, crystal structure and size of the NPs after 15 min. reaction were investigated using TEM. The synthesized nanoparticles are irregular and have faceted morphologies with a high tendency of being agglomerated (Fig. 5a). The average diameter of the NPs is 6.2 nm ± 2.5 nm (Fig. 5c). The smallest particles were only observed in HRTEM and it is likely that the poor visibility of these skews the NP size distribution to a higher value than factual. HRTEM (Fig. 5b) showed lattice fringes with an inter-planar spacing of 3.1 Å, which can be ascribed to the (112) plane of the kesterite CZTS[23]. Other lattice distances of 3.0 Å and 3.6 Å were also observed and correspond to the (013) and (111) planes of kesterite CZTS.

The XRD pattern of a drop-cast film of both samples features peaks at 2θ positions 28.7°, 47.7°, and 56.5° (Fig. 6a), which are characteristic for kesterite CZTS (JCPDS 026-0575). Significant peak broadening is observed which correlates well with the small size of NPs and the diffractogram is similar to what has previously been reported for room temperature synthesis.[11] Secondary phases such as $Cu_xS$, were not found, but the possible presence of cubic ZnS and $Cu_2SnS_3$ cannot be excluded due to the similarity of their XRD patterns with that of CZTS. Neither SnS, $SnS_2$ nor $Sn_2S_3$ could be clearly observed and the proximity of their main peaks to the CZTS-related (112)-peak hindered their identification – 32.0°, 32.2° and 33.5° for SnS, $SnS_2$ and $Sn_2S_3$, respectively. A possible peak around 52.2° matches that of wurtzite CZTS, but other wurtzite-related peaks are difficult to observe. By fitting Lorentzian curves to the peak at 47.8° 2θ we are able to apply the Scherrer equation to get an indication of crystal sizes in the batches (Fig. 6b).

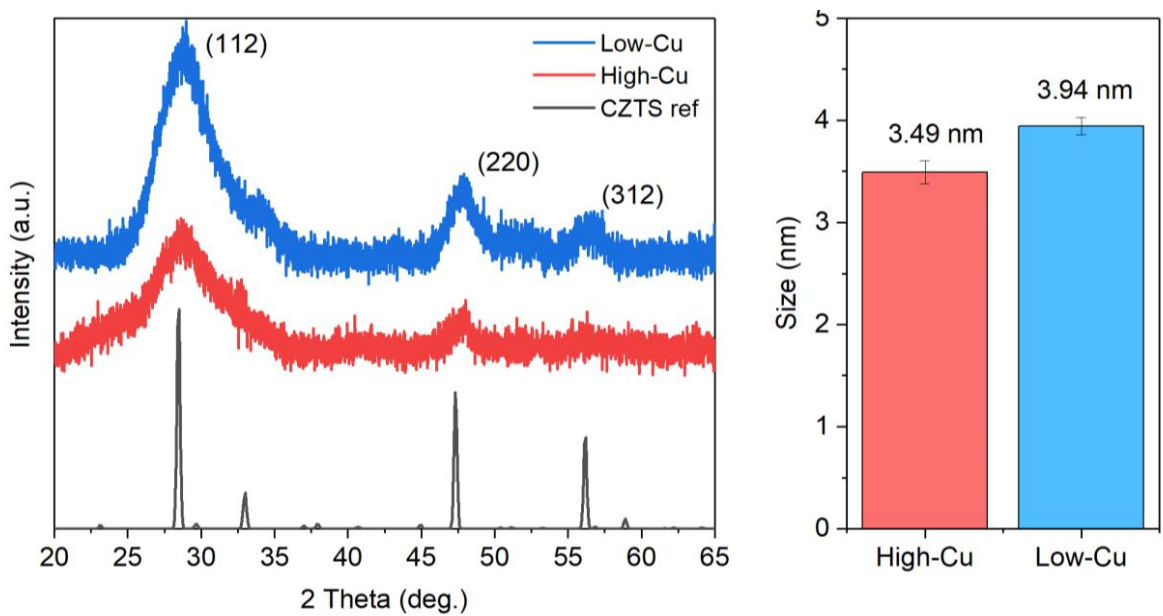

**Fig. 6** (a) XRD patterns of High-Cu (red) and Low-Cu (blue) after 15 min. reaction, as well as the reference CZTS pattern (black). (b) Average crystal size and standard deviation of each product derived from Lorentzian fits of the (220)-peak using the Scherrer equation.

The XRD data from samples with high Sn- and low Cu-content (Fig. S5, ESI) shows crystallite sizes below 3 nm, but these crystals often formed larger particles (4.3 nm ± 1.5 nm) as observed by TEM (Fig. S6, ESI). The formation of $(Sn_2S_6)^{4-}$-complexes from excess tin and sulfur as shown previously[15, 16] is likely responsible for the crystallite-clustering leading to the formation of larger particles. For samples with a reactant Cu/(Zn+Sn) ≥ 0.72, only crystallite sizes of above 3 nm are seen, with the predominant size around 4 nm (Fig. S5, ESI), with no clear correlation between reaction time and crystallite size. This particle growth with increased Cu-content could be viewed as the consumption of $(Sn_2S_6)^{4-}$-complexes for the formation of CZTS until $Cu_xS$ phases begins to dominate as observed with Raman spectroscopy. These findings are further supported by trends in band gap ($E_g$) variation (Fig. 7b).

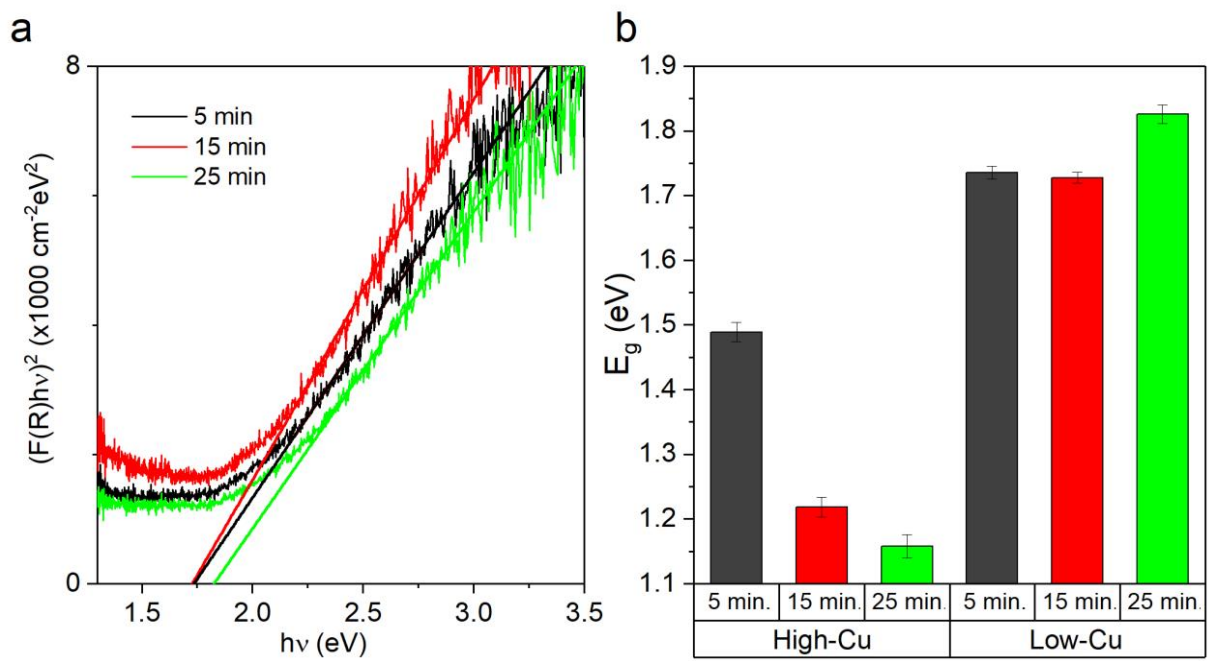

**Fig. 7** (a) Equivalent Tauc plot of UV-Vis reflectance data for Low-Cu after 5, 15 and 25 min. reaction, as well as the linear extrapolations of the band gaps. The intersection with the primary axis denotes the band gap of the NPs. (b) Band gaps derived from fitted Tauc plots.

Band gaps are derived according to the Kulbeka-Munk method from the UV-Vis reflectance curves[24] as shown for Low-Cu (Fig. 7a). First, the measured diffuse reflectance is converted to the equivalent absorption coefficient, *F(R)*. A linear extrapolation of the equivalent Tauc plot *((F(R)hv)$^2$)* versus photon energy, *hv*, is used to determine the band gap as the intersection at the primary axis. For these NPs, $E_g$ of 1.74, 1.73, 1.83 eV were obtained for products at 5 min., 15 min. and 25 min. reaction, respectively (Fig. 7b). The high $E_g$ compared to bulk CZTS is related to the expected quantum confinement in CZTS NPs that are smaller than the Bohr radius for CZTS (3.3 nm)[25] and it is possible that the effect is detectable even for particles slightly larger than the Bohr radius. In contrast, High-Cu has a significantly decreased $E_g$, which could be influenced by the presence of metallic $Cu_xS$-phases observed by Raman spectroscopy (Fig. 2). The formation of secondary phases may affect the reflectance, which could explain the High-Cu trend. The low $Cu_xS$ content in Low-Cu after 25 min. reaction time indicated by Raman spectroscopy does not appear to affect $E_g$ significantly. This further indicates the two synthesis regimes represented by two samples.

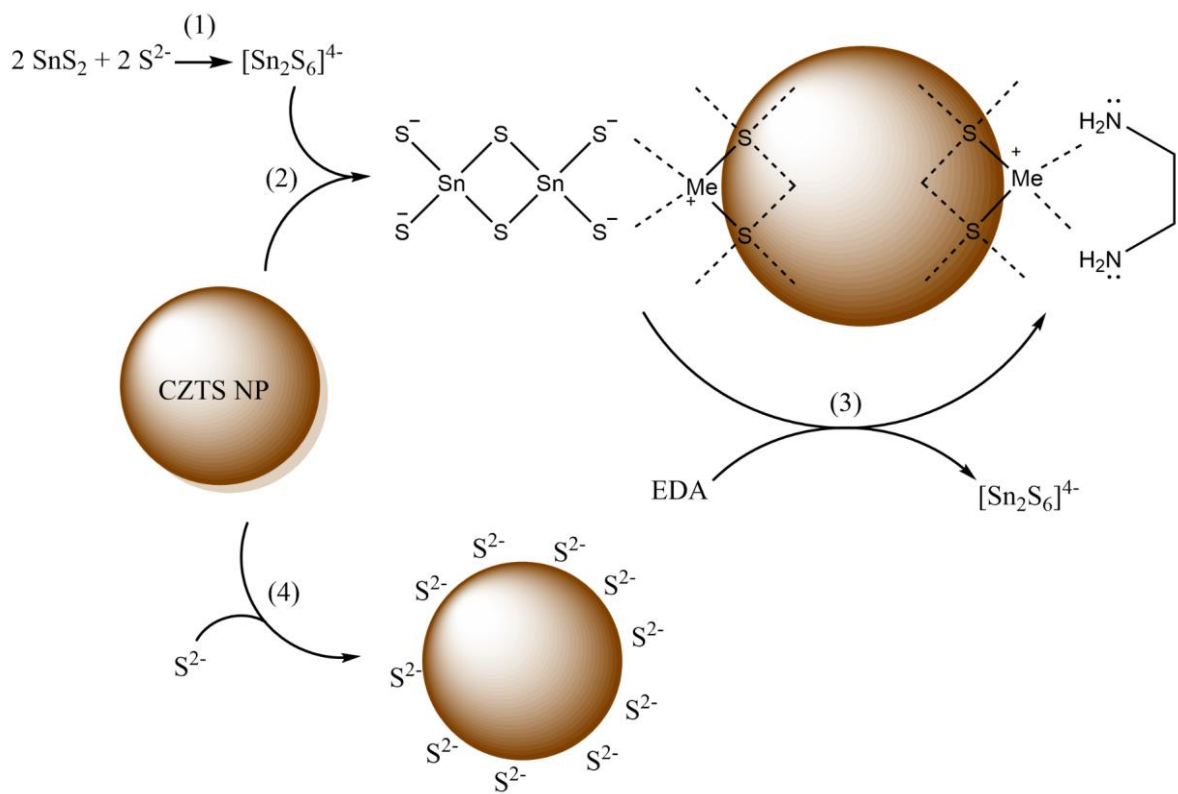

**Fig. 8** Reaction schemes for dimeric thiostannate complex formation (1), CZTS NP surface passivation using the formed complex (2), and the substitution with EDA under alkaline conditions (3). Surface passivation using sulphide ions is also illustrated (4).

The synthesized CZTS NPs, free of organic ligands, could be suspended in ethanol after sonication, but began to agglomerate and precipitate within minutes. In order to create stable inks with CZTS NPs suitable ligands and solvents are needed to promote solvent interaction with the CZTS NP surfaces. As previously described, under alkaline conditions $SnS_2$ and $S^{2-}$ forms thiostannate complexes such as $(Sn_2S_6)^{4-}$ capable of passivating the CZTS NP surface allowing us to create stable inks for solar cell printing (Fig. 8). We prepared inks using 10 mM $(Sn_2S_6)^{4-}$ in formamide, which could be destabilized upon adding ethylendiamine (EDA) (Fig. S7, ESI). The EDA is expected to substitute the complex around the NPs and promote agglomeration as a NP-to-NP linker as previously shown for other linkers, improving the solar cell efficiency up to 400%.[26]

We further examined the effect of passivating the synthesized CZTS NP surface using thiostannate complexes in a wide range of solvents. Ink stability (time before precipitation and degree of homogeneity) decreased using the following solvents (listed from highest-to-lowest stability): Formamide = DMSO > $H_2O$ > Ethanol = Acetone = Acetonitrile. Unfortunately, a concentrated ink required high amounts of thiostannate complexes, which would jeopardize the formation of a compact CZTS-layer, and the use of high boiling point solvents (formamide and DMSO) made spin coating difficult. A much more promising ink formulation is the use of small ions in alkaline aqueous solutions. Increased ink stability was found to correlate with increasing $SH^-$ and $NH_4^+$ concentrations, but for high $NH_4^+$ concentrations (10 M) the stability began to decrease (Fig. S8, ESI).

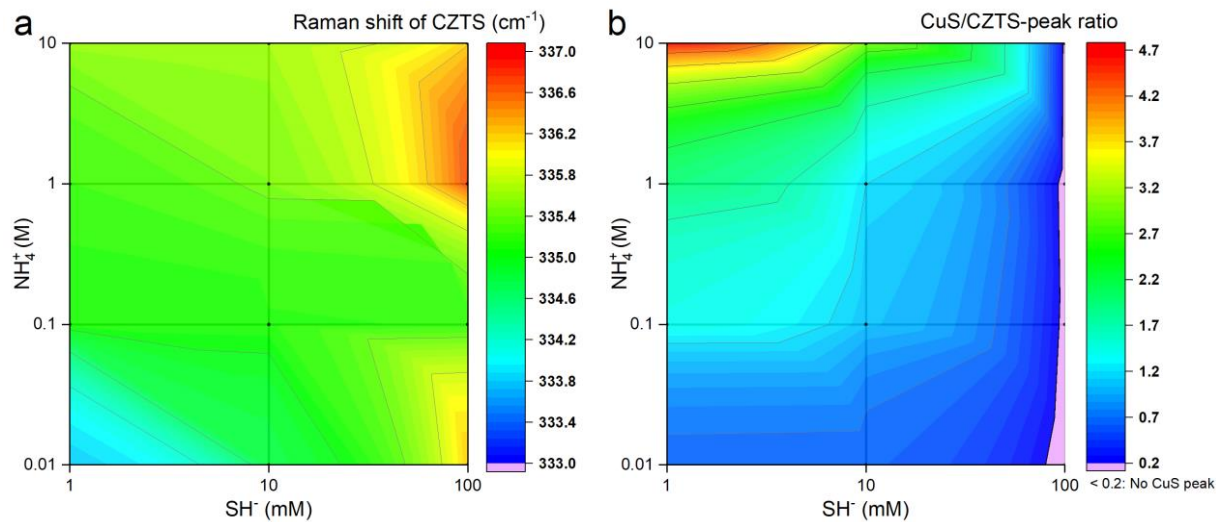

**Fig. 9** Raman spectroscopy analysis of Low-Cu CZTS NP inks with $NH_4^+$ and $SH^-$ ions for stabilization. (a) Raman shift of the main A-mode CZTS peak at different stabilizer-concentrations for identifying the intrinsic CZTS quality. (b) Relative content of $Cu_xS$ to CZTS obtained from fitted peak intensity ratios. The color gradient between points is used to guide the eye and indicate trends. No $Cu_xS$-related peaks could be clearly observed in the Raman spectra for $Cu_xS$/CZTS < 0.2 (Fig. 9S, ESI).

The formulated inks were spin-coated on glass substrates and characterized with Raman spectroscopy (Fig. 9). $NH_4^+$ and especially $SH^-$ were found to increase the Raman shift of the A-mode CZTS peak closer to 338 cm$^{-1}$ which has previously been observed for bulk kesterite-structured CZTS (Fig. 9a).[27] Compared to the low Raman shift (334 cm$^{-1}$) for samples prepared without stabilizers (Fig. 2) we assume the surface passivation is correlated to the increased Raman shift. The effect of $NH_4^+$ ions on $Cu_xS$ content diverged significantly compared to the effect of $SH^-$ ions (Fig. 9b). A direct correlation between $NH_4^+$ concentration and $Cu_xS$ was observed suggesting CZTS decomposition, whereas increasing $SH^-$ was found to suppress and even reverse the $Cu_xS$-formation. No $Cu_xS$-related peaks could be observed when using 100 mM $SH^-$ regardless of $NH_4^+$ concentration, correlating with the high ink stability (Fig. S8, ESI). Regarding CZTS decomposition, amines are known to catalyze thiostannate formation from Sn(0) by coordination and transport of tin atoms away from particle surfaces[28]. This process suppress surface passivation and the removal of tin from CZTS NPs would promote the formation of $Cu_xS$ as observed in this study for high amine concentrations.

### 4. Conclusions

Room temperature synthesis conditions of CZTS NPs were investigated for reactant ratios of Zn/Sn and Cu/(Zn+Sn) between 0.63-1.13 and 0.59-0.84, respectively. In addition, the method was improved by replacing $SnCl_4$ with the less toxic $SnCl_2$ as tin source in the synthesis. The upper limit of the reactant ratio of Cu/(Zn+Sn) was found to be 0.82 before $Cu_xS$-phases started to form at all reaction times investigated. Even at these conditions, trace amounts of $Cu_xS$-phases were identified when the reaction exceeded 15 min.

An interesting derivation from our data can be made. While XRD shows peaks that could be related to CZTS, copper-tin-sulfide (CTS) and/or ZnS, no signs of ZnS are observed in Raman spectroscopy, and EDX clearly shows high Zn-content in the nanoparticles. It is therefore likely that the nanoparticles are Zn-rich with only little or no CTS content. In addition, the absence of SnS and $Cu_xS$ (under specific conditions) is beneficial for its use as photovoltaic absorber material, and the possible presence of $SnS_2$ does not appear to significantly influence the measured band gap of the material and might help passivate the CZTS surface in the form of $(Sn_2S_6)^{4-}$-complexes. A small band gap increase indicates quantum confinement, due to the small size of the nanoparticles.

The synthesized CZTS NPs can be used for preparing highly stable carbon-free aqueous inks suitable for printing on roll-to-roll compatible surfaces for industrial scale solar cell production. The quality of the CZTS absorber layers deposited with these inks was studied with Raman spectroscopy. It showed a positive correlation between ammonia concentration and the prevalence of $Cu_xS$-phases, while the usage of 100 mM $SH^-$ (aq) was found to completely hinder the formation of any such phases.

## Associated content

*  **Supporting Information**
The Supporting Information is available:
Experimental and characterization details. Raman spectra for CZTS NPs before and after KCN etch. Size distribution of CZTS NPs. Raman peak fitting protocol.

## Author Information

**Corresponding Author**
*E-mail: chrr@dtu.dk

**Notes**

The authors declare no competing financial interest.

## Acknowledgements

This study was supported by the European Research Council (ERC) under the European Union's Horizon 2020 research and innovation programme (SEEWHI Consolidator grant, ERC-2015-CoG-681881). C. Rein gratefully acknowledges the financial support from MAX4ESSFUN of the European Regional Development Fund Interreg Öresund-Kattegat-Skagerrak (project DTU-023), Direktør Ib Henriksens Fond, and Torben og Alice Frimodts Fond. S. Engberg further acknowledges the Innovation Fund Denmark (ALTCELL project).